# Dopant redistribution and activation in Ga ion-implanted high Ge content SiGe by explosive crystallization during UV nanosecond pulsed laser annealing

Toshiyuki Tabata,[a] Karim, Huet, Fabien Rozé, and Fulvio Mazzamuto

*Laser Systems & Solutions of Europe (LASSE), 145 rue des Caboeufs, F-92230 Gennevilliers, France*

Bernard Sermage

*Probion Analysis, 37 Rue de Fontenay, F-92220 Bagneux, France*

Petros Kopalidis and Dwight Roh

*Axcelis Technologies, Inc., 108 Cherry Hill Drive, Beverly, MA 01915-1053, USA*

a) Electronic mail: toshiyuki.tabata@screen-lasse.com

## DATA AVAILABILITY STATEMENT

The data that supports the findings of this study are available within the article.

## ABSTRACT

Explosive crystallization (EC) is often observed when using nanosecond-pulsed melt laser annealing (MLA) in amorphous silicon (Si) and germanium (Ge). The solidification velocity in EC is so fast that a diffusion-less crystallization can be expected. In the contacts of advanced transistors, the active level at the metal/semiconductor Schottky interface must be very high to achieve a sub-$10^{-9}$ ohm.cm$^2$ contact resistivity, which has been already demonstrated by using the dopant surface segregation induced by MLA. However, the beneficial layer of a few nanometers at the surface may be easily consumed during subsequent contact cleaning and metallization. EC helps to address such kind of process integration issues, enabling the optimal positioning of the peak of the dopant chemical profile. However, there is a lack of experimental studies of EC in heavily-doped semiconductor materials. Furthermore, to the best of our knowledge, dopant activation by EC has never been experimentally reported. In this paper, we present dopant redistribution and activation by an EC process induced by UV nanosecond-pulsed MLA in heavily gallium (Ga) ion-implanted high Ge content SiGe. Based on the obtained results, we also highlight potential issues of integrating EC into real device fabrication processes and discuss how to manage them.

**I. INTRODUCTION**

A very specific phenomenon, so-called explosive crystallization (EC), has been well known since decades both in amorphous silicon (Si)[1-6] and germanium (Ge)[7-9], often observed when using nanosecond-pulsed melt laser annealing (MLA).[1-2,5-6,8-9] When melting such amorphous layers, as well illustrated in Ref. 9, once the formed metastable liquid part spontaneously solidifies, a fraction of the latent heat released from this solidification process is used to heat a neighboring non-molten amorphous part. If the released heat is sufficient to newly melt the underlying amorphous part, the melting front propagation continues, followed by the solidification front propagation behind. As a result, a fast travelling thin liquid layer (e.g., 27–115 Å for Si and 30–117 Å for Ge in Ref. 10) propagates through the amorphous layer.[5,9-10] Both melting and solidification front propagation velocities evolve as a function of the temperature.[3-5,8] Their experimentally estimated values are typically in the range of 5 to 20 m/s.[1-5,8] As the sub-µs timescale of nanosecond MLA process allows the process to be at thermal equilibrium[11] but chemically out-of-equilibrium,[12] homogeneous nucleation and crystal growth happens in the aforementioned thin liquid layer, resulting in polycrystalline crystal states as reported in Refs. 2, 5, and 9. In amorphous SiGe, EC can happen as in amorphous Si and Ge.[13] A molecular dynamics study predicts that the existence of dopants in Si, Ge, and SiGe alloys perturbs EC by disrupting the advancement of a planar growth front.[14]

From the viewpoint of the reduction of the access resistance to transistors, which is today dominated by the metal-semiconductor Schottky contact resistivity in advanced CMOS processes,[15] EC is an interesting approach. Recently, we have reported that the solidification front propagation velocity during nanosecond-pulsed MLA may play a dominant role in the activation of dopants beyond their equilibrium solid solubility limit in a semiconductor material such as SiGe[16] and Si.[17] Moreover, a perturbation-theory-based prediction reported for indium (In) doping in Si[18] suggests that substitutional incorporation of dopants can be improved by increasing the solidification front propagation velocity up to the EC range. Also, in the same velocity range, although some diffusion occurs in the thin moving liquid layer, the solidification process itself becomes diffusion-less[19] so that it enables to keep most of the initial dopant profile shape formed by ion implantation prior to MLA.

Nanosecond-pulsed MLA applied on gallium (Ga) ion-implanted SiGe has demonstrated an extremely low contact resistivity ($< 1 \times 10^{-9}$ ohm.cm$^2$) thanks to a strong segregation of the ion-implanted Ga atoms toward the surface.[20] From the

ACCEPTED MANUSCRIPT
ECS J. Solid State Sci. Technol. 10, 023005 (2021);
https://iopscience.iop.org/article/10.1149/2162-8777/abe2ee/meta
atomic diffusion point of view, this achievement has been obtained not by EC alone, but by the secondary-melt-induced solidification (denoted as secondary crystallization (SC) in the following discussion) which follows the EC process when enough energy is provided to the system.[9] However, a potential issue of such Ga surface segregation for actual device integration is that the beneficial initial few nanometers of the regrown highly-active Ga-doped SiGe is consumed during the following contact cleaning and metallization process steps. The EC process, as it avoids any segregation while still achieving high dopant activation, may be used as an alternative path to solve this issue since the location of the peak of active dopant profile can be tuned by adjusting the initial ion-implantation profile.

In this work, the EC process induced by nanosecond-pulsed MLA in heavily Ga ion-implanted SiGe is studied. First, the Ga ion-implantation impact on the EC process will be discussed based on the redistribution of Ga atoms. Second, the activation of the dopants by EC will be investigated by using electrochemical capacitance-voltage profiling (ECVP). To the best of our knowledge, EC activation has never been studied in the past. Finally, we will highlight potential issues of integrating EC into real device fabrication processes and discuss how to manage them.

**II. EXPERIMENTAL**

A 300mm n-type Si (100) wafer was used as a substrate. A 66-nm-thick $Si_{1-x}Ge_x$ (nominal Ge content $x$ at 0.5) layer was epitaxially grown on the substrate by the reduced pressure chemical vapor deposition technique. The deposited SiGe was partially relaxed and the macroscopic degree of strain relaxation was estimated as 30% to 40%.[21] Gallium was then ion-implanted on the SiGe epilayer at room temperature (RT) with a nominal dose of $1\times10^{16}$ at./cm$^2$ at 26 keV. The impurity projected range ($R_p$) was expected at about 20 nm in depth. For this condition, the amorphized thickness was about 55 nm (not shown, see Ref. 16). A 308 nm wavelength (i.e., UV) nanosecond pulsed excimer laser annealing (SCREEN-LASSE LT3100, more details are given in Ref. 22) was applied at RT with different laser energy densities (J/cm$^2$) and a pulse duration of 160 ns to melt the ion-implanted SiGe layer. An in-situ time-resolved reflectivity (TRR) methodology using a light of 635 nm wavelength (see Ref. 22) was used to determine EC conditions. Two representative MLA conditions were then selected for further analysis: (i) the EC with a very limited SC in terms of depth at 0.60 J/cm$^2$ [Fig. 1(a)] and (ii) the EC followed by a significant SC at 1.20 J/cm$^2$ [Fig. 1(b)]. Since the melting and solidification fronts propagation velocities of the EC process are

3SE-80-4425-L1                    SCREEN Semiconductor Solutions, Co. Ltd.

much faster than those of the SC process, the sharp peak related to the EC process appears first in TRR, then the more extended secondary peak follows it. The TRR intensity reflects the reflectivity of dynamically evolving multiple layer stack (i.e., crystal/liquid/amorphous) probed by a 635 nm wavelength light. In our experiment, the initial and final levels represent amorphous and polycrystalline SiGe, respectively. To discuss the crystalline quality, cross-sectional transmission electron microscopy (TEM) images were taken on the samples with HITACHI High-Tech H-9500 at 200 kV. The TEM specimens were fabricated by an argon (Ar) ion-milling technique. Depth profiles of Ga and Ge concentrations in SiGe were obtained by secondary ion mass spectrometry (SIMS) with ATOMIKA 4500, using $O_2^+$ ions at 1.0 kV in a detection area of $70 \times 130$ μm$^2$.

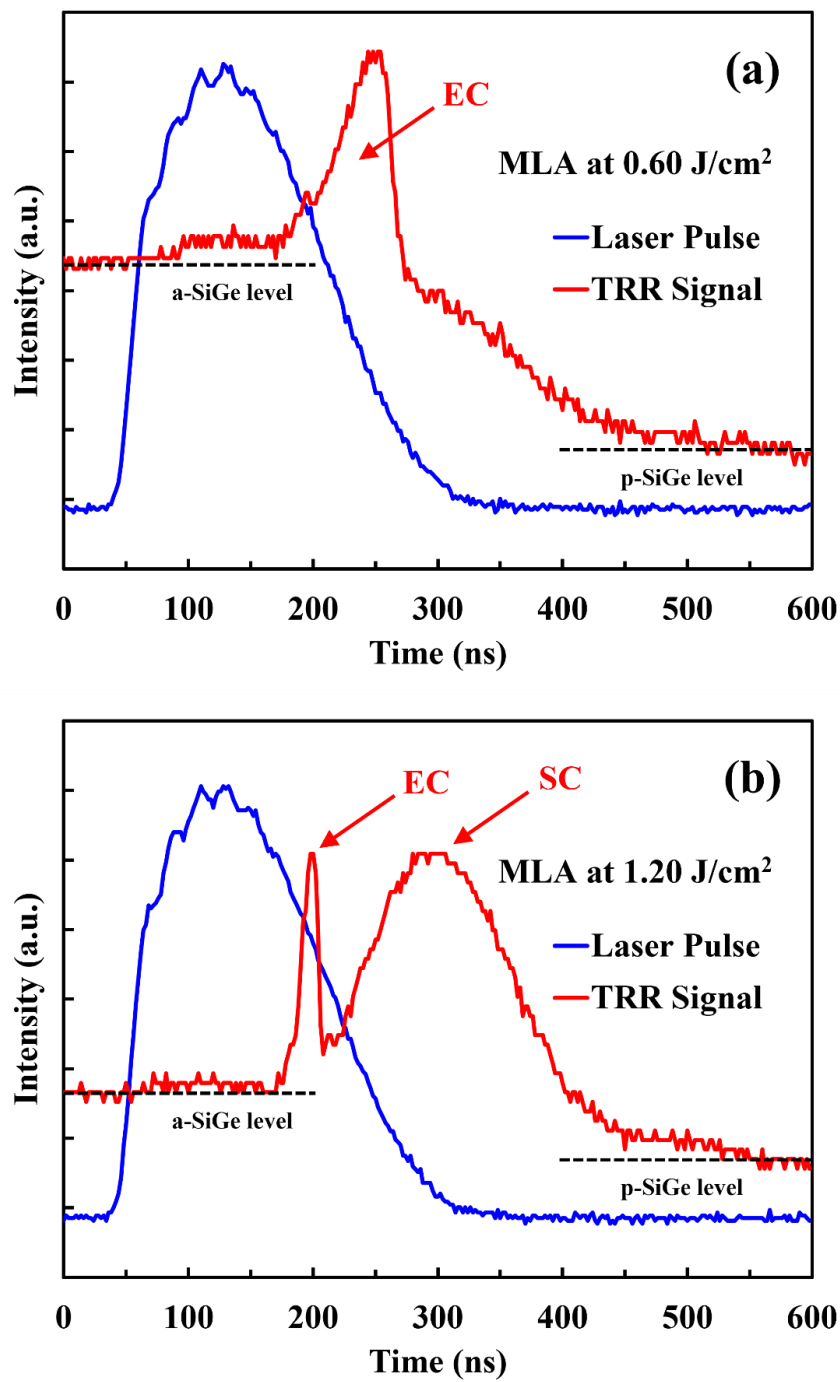

**FIG. 1.** The applied nanosecond laser pulse shape and obtained TRR signals for **(a)** the almost pure EC process at 0.60 J/cm² and **(b)** the EC process followed by the SC process at 1.20 J/cm². The parts representing the EC and SC processes are indicated by the labels "EC" and "SC" on the figures.

**III. RESULTS AND DISCUSSION**

To confirm the polycrystalline regrowth expected with the EC process, dark-field cross-sectional TEM images were taken. At 0.60 J/cm$^2$ [Fig. 2(a)], the initial 55-nm-thick amorphous SiGe layer was recrystallized into polycrystalline material down to about 30 nm in depth. The underlying layer seemingly stayed in an amorphous state. At 1.20 J/cm$^2$ [Fig. 2(b)], the initial amorphous SiGe layer was fully poly-crystallized. The Ga concentration profiles taken on the same samples are also shown, superposed to the TEM images. Interestingly, at 0.60 J/cm$^2$, a slight peak is found at the depth where the polycrystalline layer ends. Diffusion of Ga atoms is clearly observed, possibly induced by the fast-propagating EC liquid layer. According to Ref. 14, high level doping in SiGe decelerates the propagation of the thin liquid layer, where the lower limit of velocity to maintain the EC process seems to be in a range of about 4 to 8 m/s depending on the alloy composition. In this velocity range, the segregation coefficient of Ga atoms during the EC process can be still small, resulting in their incorporation into the thin liquid layer. The equilibrium segregation coefficient ($k_0$) of Ga atoms is 0.008 in Si (Ref. 23) and 0.078 in Ge (Ref. 24). The non-equilibrium segregation coefficient ($k_{ne}$) is in fact a function of the solidification front propagation velocity, and it gradually reaches unity as the solidification front propagation velocity increases.[25] The $k_0$ (or $k_{ne}$) values can be determined as $C_S/C_L = k_0$ (or $k_{ne}$), where $C_S$ and $C_L$ stand for the dopant concentrations in the solid and liquid phases at the vicinity of the moving l/s interface. At the aforementioned lower velocity limit of the EC process, one may expect that the $k_{ne}$ value of Ga atoms in SiGe is still far from unity. In Fig. 2(a), a sign of Ga segregation toward the surface can be also observed on the Ga concentration profile. It is clearly due to a very shallow SC process which was induced for this MLA condition, roughly 7 nm considering the depth at the beginning of the Ga surface segregation profile. Considering the fact that the EC process was clearly detected by TRR, although the penetration depth of a 635 nm (~1.95 eV) light in crystalline Si$_{0.5}$Ge$_{0.5}$ can be much larger than 100 nm,[26] the system would have been sensitive enough to detect the shallow SC process as well. However, the evolution of TRR signal corresponding to this SC process would have been hidden by the tail of the EC process peak signal in Fig. 1(a). The slight Ga concentration increase observed in the remaining amorphous SiGe is not considered as significant compared to the uncertainty of our Ga concentration measurement by SIMS (± 40%) and that of the sputtered depth estimation (± 10%). At 1.20 J/cm$^2$ [Fig. 2(b)], the Ga diffusion during the EC process also happened for the same reason down to the bottom of the initial amorphous SiGe layer, and the observed Ga surface segregation should be related to the SC process (roughly 15 nm in depth for its beginning, see also Ref. 16). The Ge content (% = $x \times 100$ in Si$_{1-x}$Ge$_x$, $0 \leq x \leq 1$) depth profiles were also obtained by the SIMS measurements to discuss its possible impact on the Ga surface segregation [Fig. 2(c)]. At 0.60 J/cm$^2$, the as-implanted Ge profile was not changed at all, meaning that the very shallow SC process and the EC one did not make Ge atoms diffuse. At 1.20 J/cm$^2$, the Ge content profile showed a drop of a few percent at around 20 nm in depth, while a slight Ge segregation sign was observed near the surface. It should be related to the deeper SC process, where Ge atoms could have enough time to diffuse. Indeed, the $k_0$ value of Ge atoms in SiGe alloys is reported to be 0.4,[27] being much larger than those of Ga atoms in Si (0.008)[23] and Ge (0.078).[24] From this observation, we suppose that the impact of the Ge profile modification on the surface Ga segregation is minor.

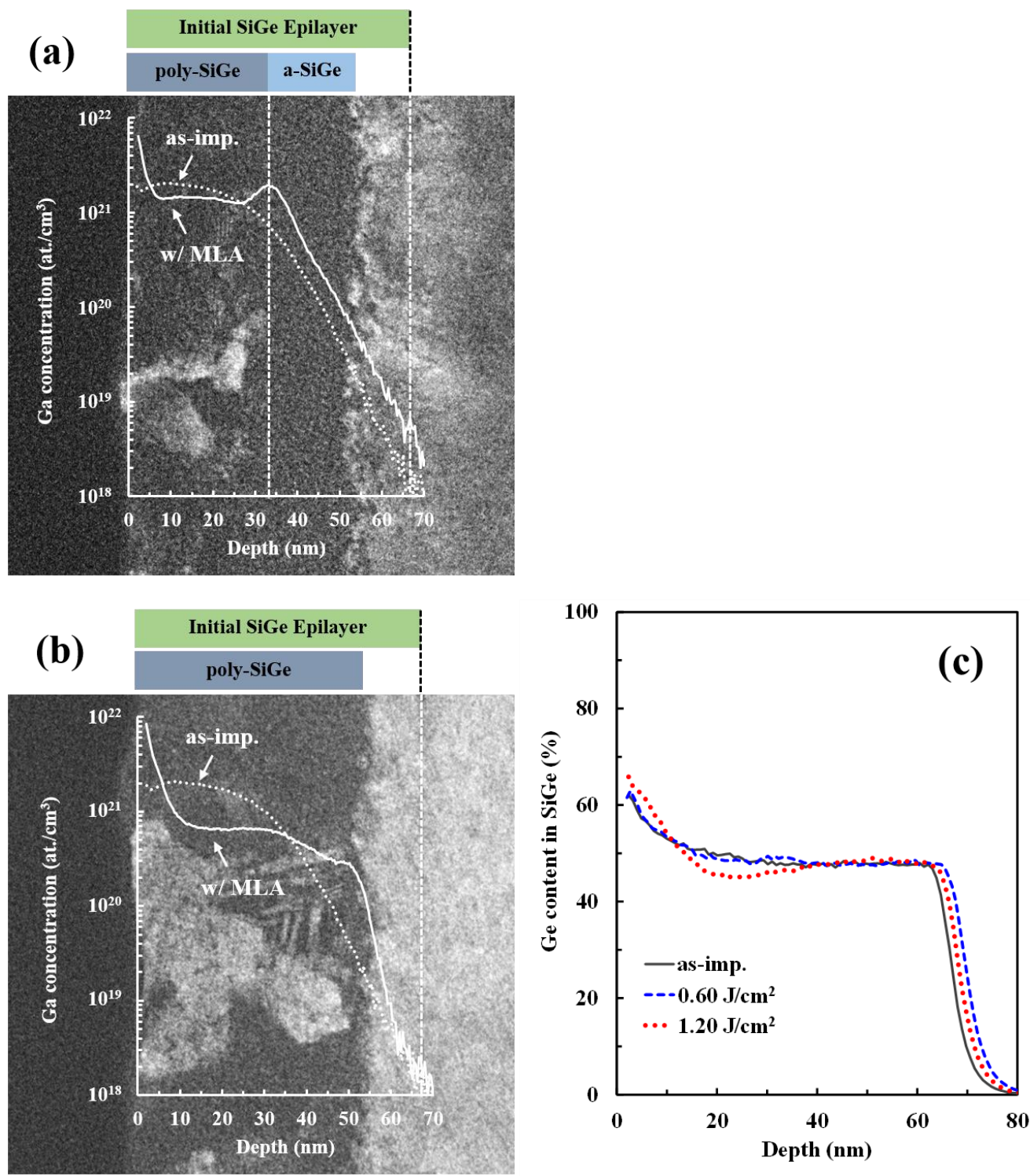

**FIG. 2.** The cross-sectional TEM images and Ga concentration depth profiles of the samples annealed at **(a)** 0.60 J/cm$^2$ and **(b)** 1.20 J/cm$^2$. The Ga concentration depth profile of the as-implanted sample is also shown. **(c)** The Ge content (% = $x \times 100$ in Si$_{1-x}$Ge$_x$, $0 \leq x \leq 1$) depth profiles obtained in the same samples.

Next, Fast Fourier Transform Mapping (FFTM) analysis was applied on the bright-field cross-sectional TEM images of the as-implanted reference and the sample annealed at 0.60 J/cm² and 1.20 J/cm² to quantitatively analyze the periodicity (i.e., grain size distribution) within the polycrystalline SiGe layer regrown by the EC and SC processes. The electron beam incident on the TEM images was on the [110] direction to the Si substrate. The local FFT spot was a circle of 1.5 nm diameter. As a result, the lower limit of detectable grain size was set to 5 nm$^2$. The analysis was carried out by moving a 10 nm-thick analysis area from the surface to the bottom of the initially amorphized layer. At a given depth, $d$, grains involved within a zone of $d$ + 10 nm were considered for the analysis (see the drawing presented as an inset of Fig. 3(a)). This is the reason why the depth range of the analysis was limited to 40 nm in depth (i.e., 40 + 10 = 50 nm, which is almost the initially amorphized thickness of 55 nm, and a margin of 5 nm was considered to avoid any perturbation from the roughness of the initial amorphous/crystalline interface). On the as-implanted reference [Fig. 3(a)], only grains slightly larger than 5 nm$^2$ were distributed over the analyzed region, well representing the amorphous feature of the as-implanted SiGe layer. On the sample annealed at 0.60 J/cm² [Fig. 3(b)], some considerably large grains up to 100 nm size could be observed. Based on the TEM and SIMS observation, the boundaries among the non-molten amorphous layer and the polycrystalline layers grown by EC and SC can be identified, respectively. The weighted average grain size is almost constant in SC, while it becomes smaller in EC and gradually reduced as the depth increases. The maximum grain size follows a similar trend. This is consistent with the EC process dynamics for which the thickness of the fast propagating liquid layer thickness is continuously reduced as it progresses towards the bottom of the amorphous layer.[9] It is also possible that the EC grains were further enlarged by the additional heat induced by the following SC process (i.e., heating of the SC liquid by laser and latent heat released in solidification). On the sample annealed at 1.20 J/cm² [Fig. 3(c)], the analyzed region is divided into two layers: the polycrystalline layers grown by EC and SC. The same observation can be made as in the case of the MLA at 0.60 J/cm². Interestingly, the gradient observed in both weighted average and maximum EC grain sizes becomes much more pronounced. The origin of this gradient could be the heat that is generated after EC by the continuous absorption of laser energy and by the larger latent heat released from the thicker SC layer. Although more sets of TEM and FFTM are required to statistically improve our analysis, the current results already evidence the qualitative difference between EC and SC in terms of the quality of regrown polycrystalline grains.

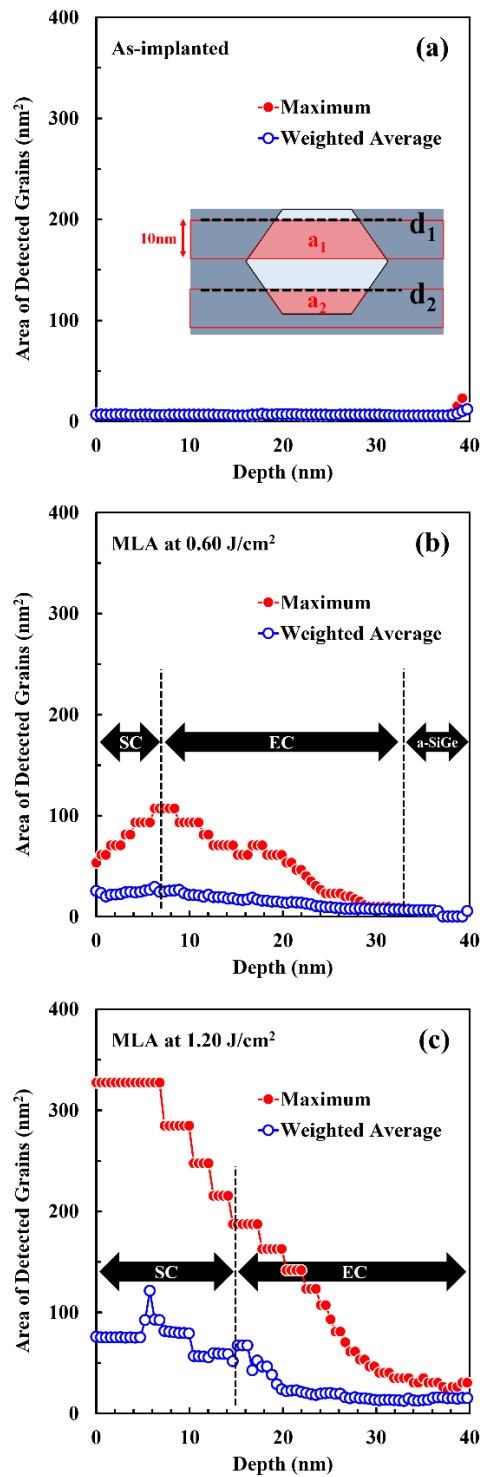

**FIG. 3.** The depth profile of the weighted average and maximum grain sizes extracted from the FFTM analysis for **(a)** the as-implanted reference and the samples annealed at **(b)** 0.60 J/cm$^2$ and **(c)** 1.20 J/cm$^2$. The parts representing the EC and SC processes are labelled as "EC" and "SC". The analysis was carried out by moving a zone of 10 nm thickness from the surface to the bottom of the initially amorphized layer. At a given depth, $d$, grains involved within a zone of $d + 10$ nm were considered for the analysis (see the drawing presented as an inset). This is the reason why the depth range of the analysis was limited to 40 nm in depth (i.e., $40 + 10 = 50$ nm, which is almost the initially amorphized

thickness of 55 nm, and a margin of 5 nm was considered to avoid any perturbation from the roughness of the initial amorphous/crystalline interface). On the inset, the area of a grain can be considered as $a_1$ at $d_1$, while $a_2$ at $d_2$. The number of effective grains (> 5 nm$^2$) involved in the analysis varied **(a)** from 1 to 6, **(b)** from 1 to 16, and **(c)** from 3 to 12 at each depth, respectively.

Finally, ECVP was carried out on the sample annealed at 1.20 J/cm$^2$ to directly compare the activation of Ga atoms in EC and SC. The details of the applied measurement and analysis method are explained elsewhere.[28] The $f(V) = 1/C^2$ curves described in Eq. 1 were measured on the sample after etching of every few nanometers. When an inverse polarization ($V$) is applied to the junction, a space charge zone is formed. The measured capacitance ($C$) is then linked to the concentration of ionized impurities ($N^*$) by the following equation.

$$\frac{1}{C^2} = \frac{2(V_{bi} - V)}{\varepsilon_r \varepsilon_0 q N^* A^2} \tag{1}$$

, where $V_{bi}$ stands for the built-in or flat band polarization, $\varepsilon_r$ and $\varepsilon_0$ for the relative and vacuum dielectric constants, $q$ for the electron charge, and $A$ for the junction area. On our sample, the $f(V) = 1/C^2$ curves showed curvatures and a discontinuity in the applied $V$ range between -1 V and 0 V. The $f(V) = 1/C^2$ curves were then fitted by a model which considered deep levels. For deep acceptors, the following phenomenological expression was used.

$$N_{da}^- = N_{da} \frac{\exp[\alpha_{da}(V_{da} - V)]}{1 + \exp[\alpha_{da}(V_{da} - V)]} \tag{2}$$

, where $N_{da}^-$ and $N_{da}$ stand for the ionized and total deep acceptor concentrations, $V_{da}$ for the polarization for which the deep acceptors were ionized, and $\alpha_{da}$ for the inverse of a potential width to consider the fact that the electric field varies inside the space charge zone. When $V > V_{da} + 1/\alpha_{da}$, $N_{da}^-$ becomes nearly equal to zero. When $V < V_{da} - 1/\alpha_{da}$, $N_{da}^-$ becomes nearly equal to $N_{da}$. The same approach can be applied also for deep donors with $N_{dd}^+$, $N_{dd}$, $V_{dd}$, and $\alpha_{dd}$.

$$N_{dd}^+ = N_{dd} \frac{\exp[\alpha_{dd}(V_{dd} - V)]}{1 + \exp[\alpha_{dd}(V_{dd} - V)]} \tag{3}$$

To calculate the $1/C^2$ value by Eq. 1, $N^*$ is defined as follows.

$$N^* = N_{sa} + N_{da}^- - N_{dd}^+ \tag{4}$$

, where $N_{sa}$ stands for the total shallow acceptor concentration. It is also possible to include different deep acceptor and donor levels in the same way. We believe that this new analysis model gives more realistic values for both shallow and deep accepter levels. In this paper, we thus focus on the evolution of the shallow ($N_{sa}$) and deep ($N_{da}$) acceptor levels with respect to the EC and SC processes. The depth profiles of the extracted $N_{sa}$ and $N_{da}$ values for the sample annealed at 1.2 J/cm² are shown in Fig. 4.

In Fig. 4, the region regrown by the SC process is from 0 nm to about 15 nm in depth, while that of EC is underneath it and goes down to the bottom of the initial amorphous SiGe layer (labelled as "SC" and "EC" in the figure, respectively). First, looking at the total concentration of the substitutional sites (i.e., $N_{sa} + N_{da}$), a maximum concentration of about $1 \times 10^{21}$ at./cm$^3$ is observed in the SC region, while it gradually decreases to about $3 \times 10^{20}$ at./cm$^3$ in the EC region for the measured range of depth. These values are higher than the equilibrium solid solubility limit of Ga atoms reported for a Ga-doped $Si_{0.4}Ge_{0.6}$ ternary alloy system (~$2 \times 10^{20}$ at./cm$^3$ from Refs. 29 and 30). This observation suggests that the substitutional incorporation of Ga atoms in SiGe strongly relies on the solidification front propagation velocity as predicted for the In-doped Si in Ref. 18, and that a slower solidification is more advantageous in our experimental setup (there should be a deviation between the case of Ref. 18 and our experiment because of the different combination of a dopant and base semiconductor material). Although additional studies are needed to quantitatively associate the ECVP analysis data with the solidification front propagation velocity at each depth, such qualitative assessment is still valid. Second, looking at the concentration of the inactive substitutional Ga atoms (i.e., $N_{da}$), a similar trend can be observed between the SC and EC regions. For instance, at 13 nm (SC) and 29 nm (EC) in depth, the measured chemical concentration of Ga atoms is almost the same, but the extracted $N_{da}$ value becomes almost two times larger in the SC region than in the EC one (i.e., ~$6.5 \times 10^{20}$ at./cm$^3$ versus ~$3.2 \times 10^{20}$ at./cm$^3$). In addition, the $N_{da}$ value increases with increase of the Ga chemical concentration and saturates around $1 \times 10^{21}$ at./cm$^3$. Third, the concentration of the active substitutional Ga atoms (i.e., $N_{sa}$) also shows a decrease from the SC region to the EC one (i.e., ~$2.5 \times 10^{20}$ at./cm$^3$ at 13 nm in depth versus ~$8.3 \times 10^{19}$ at./cm$^3$ at 29 nm in depth). One may have to remember that the presented EC region followed the additional grain growth after the EC process (see the discussion of Fig. 3(c)), which might have helped to increase the $N_{da}$ and $N_{sa}$ values with a gradient in depth. At the end, many non-substitutional Ga atoms remaining near the surface up to about 10 nm in depth of the SC region were certainly precipitated in grain boundaries.[16] A similar dopant precipitation seems to happen also in the EC region.

The high $N_{da}$ compared to $N_{sa}$ over the whole EC and SC recrystallized regions could be attributed to formation of vacancies in the Ga-doped SiGe during MLA. Indeed, in heavily arsenic (As) doped Si (i.e., roughly 1 to $2 \times 10^{20}$ at./cm$^3$ in the doped region),[31] formation of vacancies and their coupling with the substitutional dopants were suggested to explain the fact that almost half of the substitutional As atoms were electrically inactive. A similar phenomenon can be expected for our Ga-doped SiGe case, where the Ga concentration in SiGe becomes much higher than that of As in Si reported in Ref. [31]. These inactive substitutional Ga atoms could have been measured as deep acceptors in our ECVP measurement, where such deep levels are supposed to be ionized by increasing the polarization field.[28] Incorporation of points defects such as vacancies and self-interstitials during Si crystal regrowth is indeed well known and depends on the growth rate.[32] According to Ref. [33] and [34], it is also the case in Si with a nanosecond-pulsed MLA process, especially for formation of vacancies. However, their concentration is limited to a much lower level (i.e., in a range of $10^{16}$ at./cm$^3$ to $10^{17}$ at./cm$^3$) than in this work. Alternatively, by analogy with Ref. [35], another more plausible mechanism is that, when incorporating dopants into substitutional sites at an extreme dose, the crystal may lower its free energy by forming vacancies (i.e., the thermal equilibrium concentration of vacancies increases in order to compensate the strain induced by doping). In Ref. [35], a Ga doping of $\sim 1 \times 10^{21}$ at./cm$^3$ in zinc oxide (ZnO) formed Zn vacancies of $\sim 1 \times 10^{20}$ at./cm$^3$. If a $D_nV$ model (i.e., clustering of dopants (D) around a vacancy (V) with n ≥ 1) is assumed as for As atoms in Si (1 ≤ n ≤ 4),[36-38] this concentration level of vacancies has a reasonable order of magnitude with respect to the chemical Ga and $N_{da}$ levels measured in our experiment.

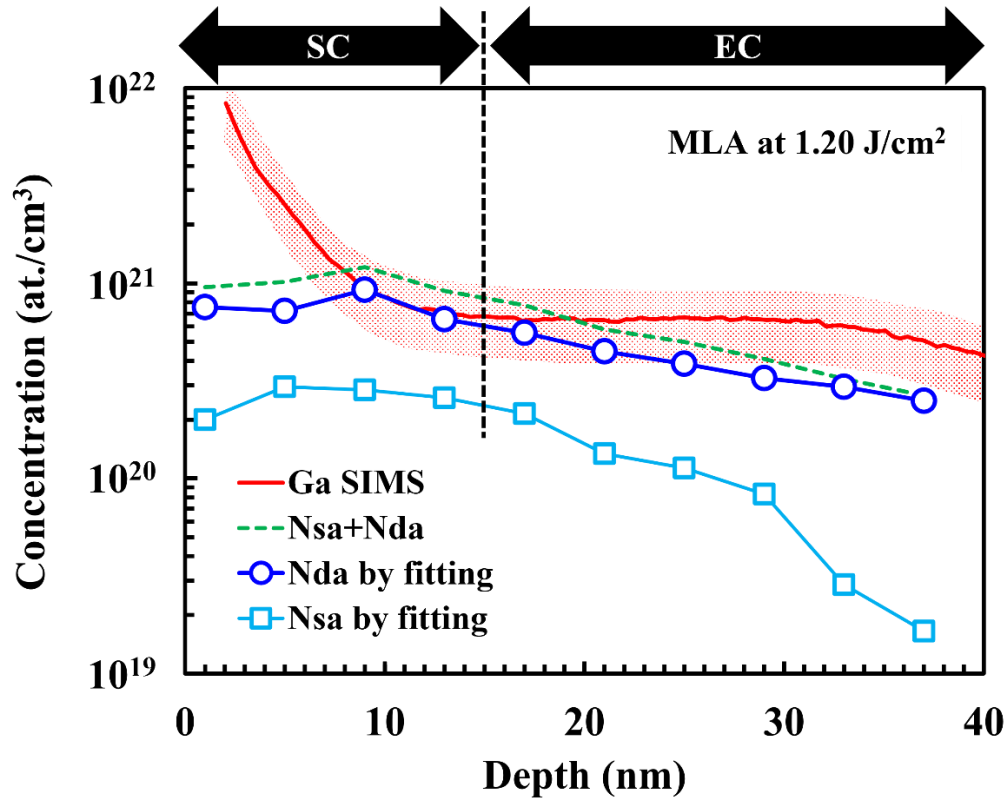

**FIG. 4.** The depth profiles of the shallow ($N_{sa}$) and deep ($N_{da}$) acceptor levels and their total ($N_{sa} + N_{da}$) extracted from our ECVP measurement and the presented $1/C^2$ fitting model (also see Ref. 28) on the sample annealed at 1.20 J/cm². The Ga SIMS profile taken at the same MLA condition is also shown with its possible error of ± 40% on the measured concentration. The parts representing the EC and SC processes are labelled as "EC" and "SC".

The results presented in this work recall some potential issues for integrating the MLA-induced EC process in real device fabrication processes. First, our results suggest that further contact resistivity lowering in future advanced transistors cannot be achieved by simply increasing the chemical concentration of dopants in semiconductor materials such as Si, Ge, and SiGe. Therefore, it is necessary to use a co-doping approach of two elements having a larger and smaller atomic radius than that of the base material, so that the formation of vacancies during MLA-induced dopant incorporation can be suppressed. In the case of a p-type co-doping in SiGe, the combination of Ga and boron (B) is a candidate.[39] An optimal mixing condition of such co-doping would strongly depend on the properties of the SiGe epilayer to be ion-implanted (e.g., the relaxation state of a SiGe epilayer is determined by the film thickness and Ge content as reported in Ref. 21). Another potential advantage of co-doping is that it can extend the thickness of the layer having the highest active dopant concentration,[39] leading to a larger process window for the subsequent surface cleaning and metallization steps. Second, as mentioned in the introduction part, an expected merit of the EC process is that it can activate dopants without significantly modifying the initial ion-implantation profile. However, our results show that significant atomic diffusion (i.e., the Ga segregation into the thin liquid layer) still happens during the EC process of heavily Ga-implanted SiGe (although it is much less significant compared to the Ga surface segregation induced by the SC process). Considering the theoretically predicted $k_{ne}$ evolution for Ga and B in Si,[25] one may expect that B is less mobile than Ga in SiGe at the same EC condition. Indeed, B does not diffuse during the EC process in Ge.[9] Since too much redistribution of dopants results in a critical reduction of the peak active level on the profile, the ion-implantation profile has to be managed so that the region to be molten by MLA becomes as shallow as possible, while keeping the margin for the subsequent surface cleaning and metallization steps. Third, the use of heavy elements like Ga for ion implantation introduces serious crystal damage in the part underlying the amorphized layer. Such damage causes the so-called End-Of-Range (EOR) defects during the annealing step and may lead to the failure of transistor operation due to a high leakage current level. To suppress such EOR defects formation, the cryogenic ion implantation can be considered as a solution.[40-41] Finally, the total dose should also be optimized to suppress the MLA-induced dopant precipitation in grain boundaries.

## IV. CONCLUSIONS

In this paper, we have investigated the dopant redistribution and activation by MLA-induced EC in the heavily Ga ion-implanted SiGe epilayer. The cross-sectional TEM and SIMS analysis revealed that Ga segregation still occurs into the thin liquid layer driving the EC process. It was supposed that the $k_{ne}$ value of Ga atoms during the EC process in SiGe was still far from unity, probably because the propagation of the thin liquid layer was perturbated by the presence of the excess Ga atoms. The FFTM analysis also supported the presence of EC and showed the structural difference of the regrown polycrystalline layers between the EC and SC cases. The ECVP analysis then suggested that most of the substitutionally incorporated Ga atoms were inactive due to the coupling with the vacancies formed during both EC and SC processes. To suppress such Ga deactivation and further improve the active level, Ga + B co-doping was suggested as a solution. From the process integration point of view, other ion implantation related conditions such as total dose, temperature, and $R_p$ were also considered to be optimized.

## ACKNOWLEDGMENTS

SL-80-4425-E1            SCREEN Semiconductor Solutions, Co. Ltd.

ECS J. Solid State Sci. Technol. 10, 023005 (2021);
https://iopscience.iop.org/article/10.1149/2162-8777/abe2ee/meta

The work realized by LASSE for this publication was supported by the IT2 project. This project has received funding from the ECSEL Joint Undertaking (JU) under grant agreement No 875999. The JU receives support from the European Union's Horizon 2020 research and innovation programme and Netherlands, Belgium, Germany, France, Austria, Hungary, United Kingdom, Romania, Israel.

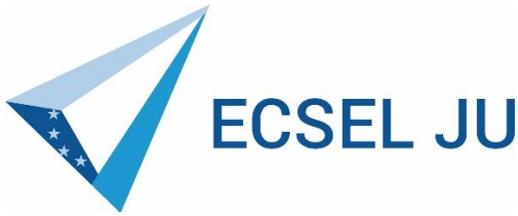
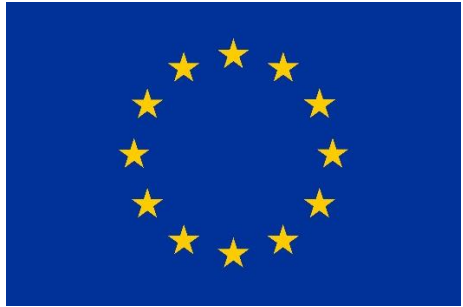

**FIGURES**

**FIG. 1.** The applied nanosecond laser pulse shape and obtained TRR signals for **(a)** the almost pure EC process at 0.60 J/cm$^2$ and **(b)** the EC process followed by the SC process at 1.20 J/cm$^2$. The parts representing the EC and SC processes are indicated by the labels "EC" and "SC" on the figures.

**FIG. 2.** The cross-sectional TEM images and Ga concentration depth profiles of the samples annealed at **(a)** 0.60 J/cm$^2$ and **(b)** 1.20 J/cm$^2$. The Ga concentration depth profile of the as-implanted sample is also shown. **(c)** The Ge content (% = $x \times 100$ in Si$_{1-x}$Ge$_x$, $0 \leq x \leq 1$) depth profiles obtained in the same samples.

**FIG. 3.** The depth profile of the weighted average and maximum grain sizes extracted from the FFTM analysis for **(a)** the as-implanted reference and the samples annealed at **(b)** 0.60 J/cm$^2$ and **(c)** 1.20 J/cm$^2$. The parts representing the EC and SC processes are labelled as "EC" and "SC". The analysis was carried out by moving a zone of 10 nm thickness from the surface to the bottom of the initially amorphized layer. At a given depth, $d$, grains involved within a zone of $d + 10$ nm were considered for the analysis (see the drawing presented as an inset). This is the reason why the depth range of the analysis was limited to 40 nm in depth (i.e., 40 + 10 = 50 nm, which is almost the initially amorphized thickness of 55 nm, and a margin of 5 nm was considered to avoid any perturbation from the roughness of the initial amorphous/crystalline interface). On the inset, the area of a grain can be considered as $a_1$ at $d_1$, while $a_2$ at $d_2$. The number of effective grains (> 5 nm$^2$) involved in the analysis varied **(a)** from 1 to 6, **(b)** from 1 to 16, and **(c)** from 3 to 12 at each depth, respectively.

**FIG. 4.** The depth profiles of the shallow ($N_{sa}$) and deep ($N_{da}$) acceptor levels and their total ($N_{sa} + N_{da}$) extracted from our ECVP measurement and the presented $1/C^2$ fitting model (also see Ref. 28) on the sample annealed at 1.20 J/cm$^2$. The Ga SIMS profile taken at the same MLA condition is also shown with its possible error of ± 40% on the measured concentration. The parts representing the EC and SC processes are labelled as "EC" and "SC".